\documentclass[11pt, a4paper]{article}
\usepackage[left=2.5cm, right=2.5cm, top=3.3cm, bottom=2.7cm, 
head=30pt]{geometry} 
\usepackage{titlesec}
\usepackage{graphicx, xcolor} 
\usepackage{fancyhdr}
\usepackage{lipsum, blindtext}
\usepackage{ragged2e, lmodern, times}
\usepackage[english]{babel}
\usepackage{siunitx}
\usepackage[utf8]{inputenc}
\usepackage[export]{adjustbox} 
\usepackage{multicol,bm,hyperref,subcaption,appendix,listings,array,url, parskip,float,csquotes,tikz,caption,booktabs,amsmath,amsthm,amssymb,enumitem,chngcntr}
\usepackage[normalem]{ulem}
\usepackage{listings}
\usepackage{textcomp, gensymb}
\usepackage{longtable}
\usepackage[export]{adjustbox}
\newcommand{\quo}{`}

\makeatletter
\renewcommand*\env@matrix[1][\arraystretch]{%
  \edef\arraystretch{#1}%
  \hskip -\arraycolsep
  \let\@ifnextchar\new@ifnextchar
  \array{*\c@MaxMatrixCols c}}
\makeatother

\usepackage{cleveref} 
\titleformat*{\section}{\fontsize{11}{11}\bfseries}
\titleformat*{\subsection}{\fontsize{11}{11}\itshape}
\titleformat*{\subsubsection}{\fontsize{14}{14}\bfseries}
\titlelabel{\thetitle.\: } 
\titlespacing*{\section}{0pt}{1\baselineskip}{0\baselineskip} 
\titlespacing*{\subsection}{0pt}{1\baselineskip}{0\baselineskip} 
\titleformat*{\paragraph}{\fontsize{14}{14}\bfseries}
\titleformat*{\subparagraph}{\fontsize{14}{14}\bfseries}

\begin{document}

\renewcommand{\headrulewidth}{0pt} 

\vspace*{-1mm}
\fontsize{17}{17}\selectfont 
\textbf{Hard constraint projection in a physics informed neural\\ network} 
\vspace{5mm}

\setlength{\leftskip}{25mm}
\fontsize{10}{10}\selectfont 
\textbf{
M J S Horne$^{*}$$^{1}$,
P K Jimack$^{1}$,
A Khan$^{2}$,
H Wang$^{3}$
}\par
\fontsize{9}{9}\selectfont 
$^{1}$ 
School of Computer Science, University of Leeds, Leeds, LS2 9JT, UK\\
$^{2}$
School of Civil Engineering, University of Leeds, Leeds, LS2 9JT, UK\\
$^{3}$
UCL Centre for Artificial Intelligence, Department of Computer Science, University College London, London, WC1E 6BT, UK\par
\setlength{\leftskip}{0pt}

\vspace{4mm}
\setlength{\leftskip}{25mm}

\fontsize{10}{10}\selectfont 
\par
\justifying
\fontsize{10}{10}\selectfont
\setlength{\leftskip}{25mm}
\noindent
\textbf{Abstract: }
Physics informed machine learning provides a framework to study and simulate fluid dynamics, potentially at a fraction of the computational cost of traditional numerical methods, with the goal of improving the explainability, accuracy and robustness of purely data driven machine learning methods.
In this work, we extend the hard constraint projection (HCP) method developed by Chen \textit{et al.}
[Journal of Computational Physics, 445,  110624, 2021] 
from a linear partial differential equation (PDE) to a nonlinear PDE: the 1D Burgers' equation.
To strictly impose a discrete approximation of the governing equation, the Burgers' equation is first locally discretised (using a finite difference stencil) then decomposed into a constraint matrix and a prediction matrix. Using tools from linear algebra, the constraint matrix is used
to construct a projection matrix.
A physics informed neural network (PINN) 
[Raissi \textit{\textit{et al.}} Journal of Computational Physics, 378, 686 – 707, 2019]
is used to predict the velocity,
and a final unlearnable layer projects the velocity prediction to the hyperplane of exact solutions to a discretised form of the Burgers' equation.
Our initial results indicate that the HCP-PINN initially trains similarly to an equivalent PINN, and as the PINN loss plateaus, the HCP-PINN loss continues to decline.
When tested on a previously unseen uniform grid of points covering the domain, the HCP-PINN exhibits higher accuracy, converging to a data error approximately an order of magnitude lower than the comparable PINN. We also observe that the inclusion of the HCP layer results in better observance of the governing PDE.
Our findings suggest that the HCP-PINN can be more robust to the selection of physics collocation points, and less dependent on collocation point quantity.


\par
\vspace{4mm}

\setlength{\leftskip}{0pt} 


\section{Introduction}
\fontsize{11}{11}\selectfont
The dynamics of fluids are governed by nonlinear partial differential equations (PDEs), the solution of which remains an open field of research, even after centuries of study.
Utilising recent developments in computational power, along with decades of simulated computational fluid dynamics data, machine learning is a promising framework to simulate fluid dynamics at a fraction of the computational cost of traditional numerical methods\cite{brunton2020rev}. 
With {the influx of available} computationally simulated or experimentally gathered data, one machine learning approach is to feed this data into a neural network to learn the evolution of the system entirely from the data\cite{TONG2020gappypod,NAKAMURA2021CNNAELSTM}. The model can then make predictions for similar but unseen scenarios faster than the equivalent numerical simulations. 

\noindent In the setting where we are seeking solutions to a known governing equation, the purely data-driven models neglect this physical knowledge.
In 2019, Raissi \textit{et al.} \cite{raissi2019ogPINN} laid the foundations for the field of physics informed machine learning with their novel architecture, the physics informed neural network (PINN). 
The innovation of the PINN lies in the formulation of the loss function, which is comprised of two weighted and summed elements, the data error (which may itself be a weighted sum of the boundary and internal data error) and the physics error (the residual of the governing equation, calculated using automatic differentiation on the model prediction, with respect to the input coordinates).
It has been reported that the predicted solutions generated by a PINN produce more accurate results than the data driven models\cite{raissi2019ogPINN}, while being more robust to noisy\cite{shone2023cardiacmriPINN} and sparse\cite{cai2021espresso} training data. 
The interpretability of the governing equation residual provides a more explainable network, an improvement over the black box learning of purely data driven networks.

\noindent However, the introduction of a physics based term in the loss function is not without its challenges, and in some settings the training of the PINN will fail to converge\cite{bajaj2021G-PINN}.
By examining the loss landscape, Krishnapriyan \textit{et al.} \cite{krishnapriyan2021PINNflaws} demonstrated that the inclusion of the physics term in the PINN loss causes the optimisation problem to become worse conditioned, with more local minima. Given that nonlinear PDEs have complex solution spaces, sensitive to initial conditions and PDE coefficients, Krishnapriyan \textit{et al.} \cite{krishnapriyan2021PINNflaws} also identified that the loss landscape became more complex with more difficult parameter regimes (for example, in fluids with high Reynolds numbers).

\noindent As the evolution of a fluid is partially determined by the initial and boundary conditions, efforts have been made to build PINNs which must obey these strictly.
Mohan \textit{et al.} \cite{mohan2020PhyCAE} and Chen \textit{et al.} \cite{chen2021HCP} both employ ghost cells (virtual cells outside the domain), a method from the field of computational fluid dynamics, to discretely enforce the required initial and boundary conditions.
Lu \textit{et al.} \cite{lu2021hardcon_PINN_inverse} strictly enforce both Dirichlet and periodic boundary conditions by making an intermediate prediction which is then passed through a user defined function that satisfies the condition by construction. 

\noindent Some authors have found that strictly enforcing a subset of the governing equations generates more favourable results than weakly enforcing the full set. 
Mohan \textit{et al.} \cite{mohan2020PhyCAE} and Raissi \textit{et al.} \cite{raissi2019ogPINN} both enforced incompressibility in their fluid by making an initial prediction that could be transformed using methods from vector calculus to ensure a divergence free flow 
(the Helmholtz decomposition\cite{mohan2020PhyCAE} and the stream function\cite{raissi2019ogPINN}).

\noindent Beucler \textit{et al.} \cite{BEUCLER_TOM2021analconst} transformed the system constraints into a system of linear equations, with nonlinear variables in the case of nonlinear constraints.
Their network predicts a subset of the variables, and then uses the residual of their linear system to define the remaining variables. In this way the full set of variables fulfil the constraints exactly, however the authors report that the residual-determined variables had consistently higher data errors than the predicted variables.
N\'egiar \textit{et al.} \cite{negiar2023hardconst} also transform the system constraints into a system of equations to be solved. They use a neural operator to predict a set of $N$ 1D functions which are combined in a weighted sum to satisfy the physics on the $n$ training coordinates. The system of $n$ constraints and $N$ variables can be solved to find the weights of the summation.
If the constraints are nonlinear, then a nonlinear least-squares method is required to find the weights, and if the system is overdetermined ($N<n$) the user must choose between satisfying the constraints on a subset of the data or only satisfying the system approximately.

\noindent Chen \textit{et al.} \cite{chen2021HCP} strictly embedded their governing linear PDE by defining a projection that pulls an arbitrary prediction to the closest point on the hyperplane of exact solutions to the discretised form of their governing equation. When Chen \textit{et al.} tested this method on ground water flow through a porous medium, they reported that during training all components of the loss function declined faster, converging in fewer epochs than a comparable PINN.

In this paper, we extend the method of hard constraint projection used by Chen \textit{et al.} \cite{chen2021HCP} from a linear PDE to a nonlinear PDE, the 1D Burgers' equation, and start to investigate the effect of the projection on the training of a neural network.


\section{Physics Informed Neural Network}
We will consider a nonlinear time dependant PDE of the form
    ${\partial 
    \boldsymbol{u}}/
    {\partial t} = 
    \mathcal{N}
    (\boldsymbol{x}, 
    \boldsymbol{u}, 
    \nabla \boldsymbol{u}, ...)$
where the nonlinear operator $\mathcal{N}$ is known.
We are seeking the solution $\boldsymbol{u}$ at time and space coordinates 
$(t,\boldsymbol{x})$ 
subject to initial and boundary conditions.
{A feed forward neural network (FFNN) with learnable parameters $\boldsymbol{\theta}$ will be used to approximate our PDE solution 
$\mathrm{FFNN}_{\boldsymbol{\theta}}(t_i,\boldsymbol{x}_i)=
\boldsymbol{\hat{u}}_i(t_i,\boldsymbol{x}_i)$}.

\noindent 
Given ground truth data of our PDE solution  
$( \boldsymbol{u}_i(t_i,\boldsymbol{x}_i),i\in \{1,2,..,N\} )$, on the spatio-temporal boundaries and/or within the domain, the loss function can be defined as the distance between the ground truth data and the network's predictions at those coordinates. We will use the mean squared error as the distance measure. 
The data error based loss is defined on the ground truth data set, see equation (\ref{loss_de}).
\begin{equation}
\label{loss_de}
\mathcal{L}=
\mathrm{DE}= 
\frac{1}{N}\sum_{i=1}^{N}
(||
\boldsymbol{u}_i(t_i,\boldsymbol{x}_i)
-
\boldsymbol{\hat{u}}_i(t_i,\boldsymbol{x}_i)
||)^2
\end{equation}
Training the model only on the data error (DE) neglects the other information we have on the system, the governing PDE. 
A PINN (as proposed by Raissi \textit{et al.} \cite{raissi2019ogPINN}) appends the data-driven loss function with a measure of the prediction's deviations from the governing equation, the physics error (PE). 
\begin{equation}
    \label{gen_res_PE}
    \mathrm{PE}
    =
    \frac{1}{M}\sum_{i=1}^{M}
    (||
    \boldsymbol{r}_i(t_i,\boldsymbol{x}_i)
    ||)^2,
  \;\;\;\;\;\;\;\;\;
    \boldsymbol{r}
    =
    \frac{
    \partial 
    \boldsymbol{\hat{u}}}{
    {\partial t}} - 
    \mathcal{N}
    (\boldsymbol{x}, 
    {\boldsymbol{\hat{u}}}, 
    \nabla \boldsymbol{\hat{u}}, ...)
\end{equation}
\noindent We define the governing equation residual as $\boldsymbol{r}$ in equation (\ref{gen_res_PE}), which will be zero if the known physics is obeyed exactly.
We choose $M$ points in the domain to evaluate the residual at, and the physics error is defined as the mean square of these residuals (PE, (\ref{gen_res_PE})). 
The partial derivatives required to calculate the residual are found using automatic differentiation.
Automatic differentiation exploits the differentiable chain of linear sums and nonlinear activation functions to construct exact partial derivatives with respect to the input variables.
\begin{equation}
\label{loss_de_pe}
\mathcal{L}=
(w)\mathrm{DE} +
(1- w)\mathrm{PE}
\end{equation}
The PINN loss function is defined by equation (\ref{loss_de_pe}), where $w$ is a (user-defined) hyperparameter that controls the weighting of the two components.

\section{Hard Constraint Projection}
\begin{figure}[t] 
\centering
\includegraphics[height=4cm]{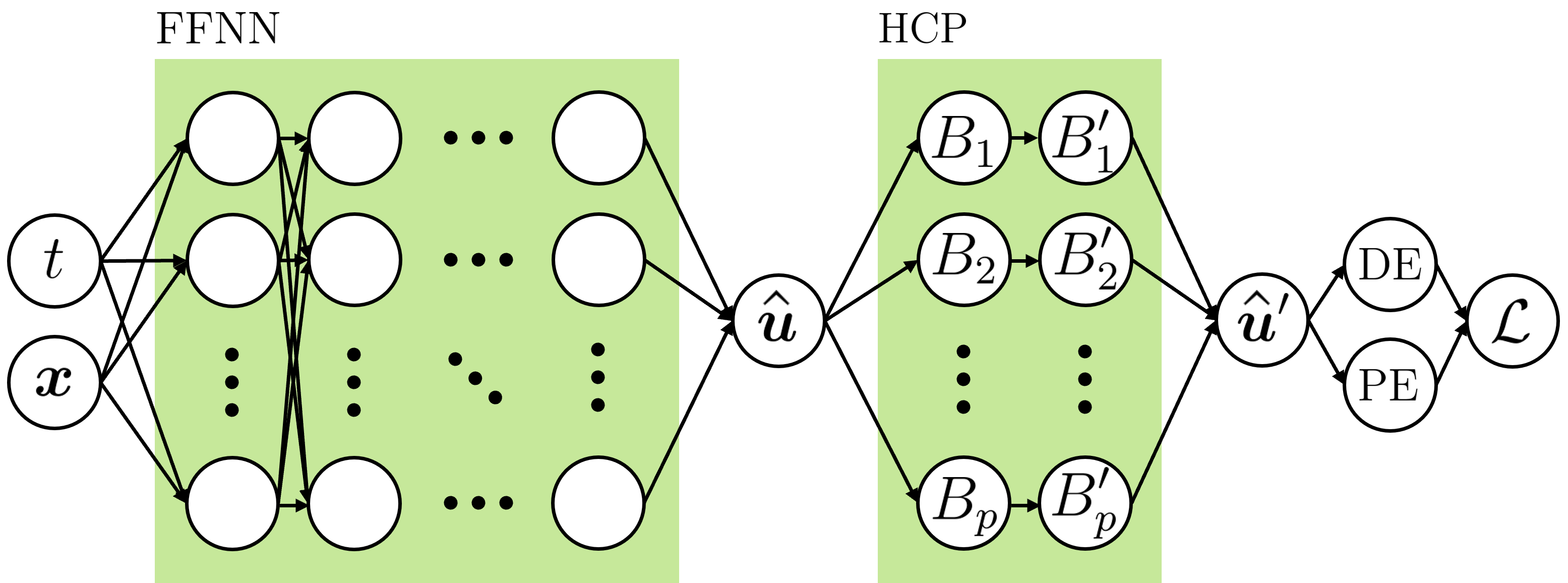}
\caption{{HCP-PINN} architecture. The left box is the {FFNN} with learnable weights and biases, and the right box is the unlearnable {HCP}.}
\label{41archi_hcp}
\end{figure}

\noindent 
The HCP-PINN has the same learnable network and loss function as a PINN, but between the initial prediction of the solution and the loss evaluation there is an unlearnable hard constraint projection layer (see figure \ref{41archi_hcp}).
The implementation of this method has three offline steps. 
First, the governing PDE (in this work, the Burgers' equation, (\ref{burgers1d})) is discretised with a finite difference approximation for the derivatives.
Following the methodology used by Chen \textit{et al.} \cite{chen2021HCP}, we choose a central finite difference scheme for the spatial derivatives, and a backwards finite difference scheme for the temporal derivative.

\noindent
Second, the discretised form of the governing equation is decomposed into two matrices, the constraint matrix $A$ (containing all the constant terms) and the prediction matrix $B$ (containing all of the transient terms) such that multiplying $AB$ recovers the discretised governing equation (see equation (\ref{burgers_discA})).
\begin{equation}
    \frac{\partial u}{\partial t} +
    u \frac{\partial u}{\partial x}
    - \nu
    \frac{{\partial^2 u}}{\partial x^2}
    =0.
    \label{burgers1d}
\end{equation}
\begin{equation}
    AB = 
    \begin{bmatrix}
        \frac{1}{\Delta t} &
        \frac{1}{\Delta t} &
        \frac{1}{\Delta x} &
        \frac{1}{(\Delta x)^2}
    \end{bmatrix}
    \begin{bmatrix}[1.32]
        u(t,x) \\
        -u(t-\Delta t,x) \\
        \frac{1}{2}
            u(t,x)[u(t,x+\Delta x) - 
            u(t,x-\Delta x)] \\
        -\nu 
            [u(t,x+\Delta x) 
            -2u(t,x) + 
            u(t,x-\Delta x)]
    \end{bmatrix}
    \label{burgers_discA}
\end{equation}

\noindent
The third step is to define a projection matrix to be used during model evaluation to enforce our governing equation. 
Using tools from linear algebra, we can define the projection matrix as $ P = I - A^T(AA^T)^{-1}A$. 
When an arbitrary prediction matrix ($B$) is multiplied by the projection matrix, $PB=B'$, the outcome ($B'$) is the closest point to $B$ that lies on the hyperplane $AB=0$. 
For a proof of this, please see the appendix of the paper by Chen \textit{et al.} \cite{chen2021HCP}.

\noindent
During training (or testing) the input to our network is sets of coordinates $(t_i,x_i),\: i\in\{1,2,..,p\}$. 
The model predicts the solution at these points $\hat{u}(t_i,x_i)$. 
We also evaluate the model at the nearby points required to construct the chosen discretisation stencil (here: 
$\hat{u}(t_i-\Delta t,x_i),
\hat{u}(t_i,x_i+\Delta x),
\hat{u}(t_i,x_i-\Delta x)$).
With these predictions, the prediction matrix $B(t_i,x_i)$ can be constructed, and consequently projected using the projection matrix we constructed offline, $B'(t_i,x_i) = PB(t_i,x_i)$. 
The HCP-PINN prediction $\hat{u}'(t_i,x_i)$ can be extracted from $B'(t_i,x_i)$ for a solution that satisfies the discretised governing equation exactly, locally.
During training the loss is evaluated on our projected prediction $\hat{u}'(t_i,x_i)$.



\section{Results}
To test the hard constraint projection, we train a PINN with and without the HCP layer, with all other hyperparameters fixed.
The networks both have 5 hidden layers of 50 neurons each, with the hyperbolic tangent (tanh) as the activation function.
Adam optimiser\cite{kingma2017adam} with default settings is used to train our networks, and the weight in our loss function ($w$ in equation (\ref{loss_de_pe})) is set to $0.9$.
The discretisation used in the projection 
($\Delta t, \Delta x$ in equation (\ref{burgers_discA})) is set to 
{($10^{-9},10^{-9}$)}.

\begin{figure}[t] 
\centering
\includegraphics[width=\linewidth]{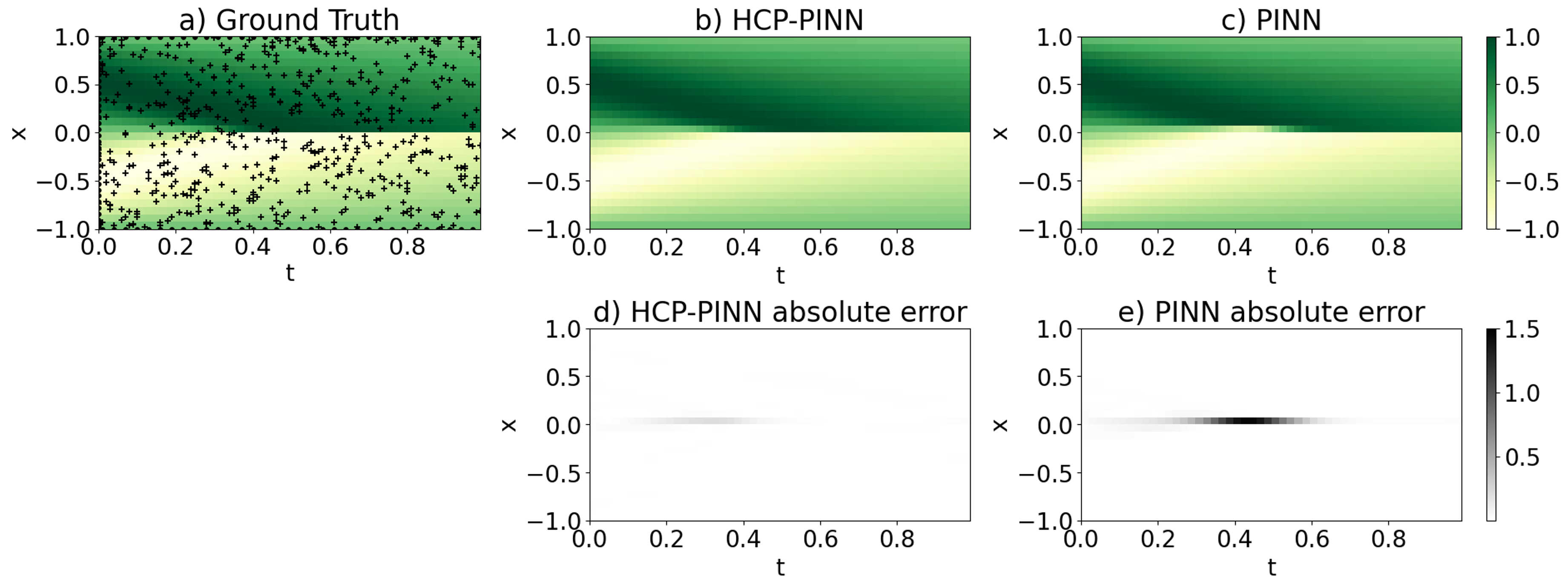}
\caption{
(a) Foreground: the sampling strategy of the training data, with $\bullet$ for the locations of the data error calculations, and $+$ for the locations of the physics error calculations. Background: ground truth associated with the test data set. 
(b),(c) Predicted solution at the end of training, on the test set. 
(d),(e) Absolute error in solution predictions. 
Colour bars shared by plots in the same rows.}
\label{51_qualitative}
\end{figure}

\noindent
The data used to train our models is 
a single 1D Burgers' equation (\ref{burgers1d}), with domain $(t,x) \in (0,1) \times (-1,1)$, 
with a sine wave initial condition ($t=0$), Dirichlet boundary conditions ($u=0$; $x=-1,1$), and viscosity $\nu = 0.01/\pi$. The training dataset is split into ground truth values (on which the DE is calculated) and physics collocation values (on which the PE is calculated). 
The ground truth values are provided only for the initial and boundary conditions, with 32 at $t=0$ and $2\times33$ at $x=-1,1$, a total of $N=98$, which are represented by $\bullet$ in figure \ref{51_qualitative}a.
The physics collocation points are random uniformly sampled throughout the interior of the domain, with $M=500$, represented by $+$ in figure \ref{51_qualitative}a.
The test loss is composed of a DE and PE calculated on the same coordinate set ($N_{\mathrm{test}}=M_{\mathrm{test}}=1300=26\times50$), a grid of points blanketing the entire domain. The grid of test coordinates can be observed by the \quo pixels' in figure \ref{51_qualitative}.

\begin{figure}[t] 
\centering
\includegraphics[width=\linewidth]{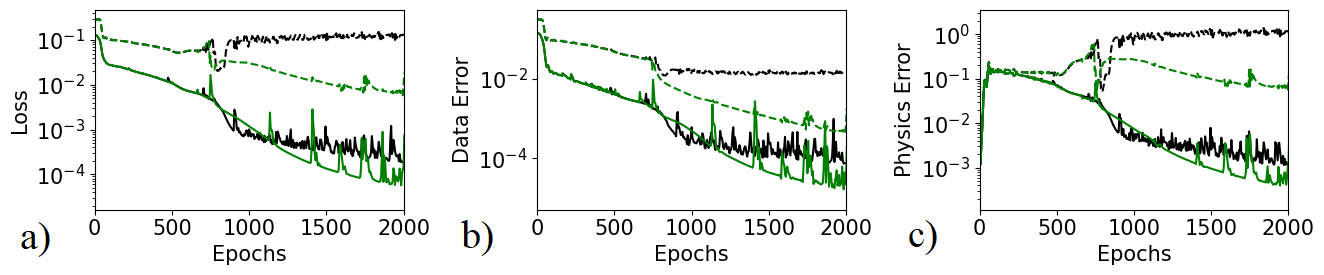}
\caption{Graphs comparing the value of the (a) loss, (b) data error, and (c) physics error, on the test and training set, during the training of both models. The black lines relate to the PINN, and green lines the HCP-PINN. The solid line is the metric evaluated on the training data, and dashed line on the test data.}
\label{52_quantative}
\end{figure}

\noindent
In figure \ref{52_quantative} we have provided the value of the loss, data error, and physics error of the two models during training, evaluated on the training dataset (solid line) and the test dataset (dashed line).
We can see that the PINN and HCP-PINN train similarly for the first $\sim$700 epochs, and that around the 1000th epoch the PINN seems to plateau, with the HCP PINN continuing to train, with both components of the loss declining. 
It appears that the PINN is overfitting after epoch 1000, as the training metrics continue to fall without improvement to the test metrics.
The value of the test physics error is lower for the HCP-PINN than for the PINN (see the dashed lines in figure \ref{52_quantative}c), which we would expect from a model where the governing equations are strictly imposed.
We would not expect the physics error of the HCP-PINN to be exactly zero, as the governing equation is only exactly obeyed in its discretised form, locally.
The test data error measures how closely the predicted solution matches the ground truth inside the domain. We can see that at the end of training the test DE of the  HCP-PINN is over an order of magnitude lower than the test data for the PINN, evidence that the inclusion of the hard physical constraint can improve the prediction accuracy along with enforcing the underlying physics.

\noindent
For a qualitative analysis we can compare the predictions made by the HCP-PINN (figure \ref{51_qualitative}b) to those made by the PINN (figure \ref{51_qualitative}c). Neither model has access to solution data inside the domain, only information on the physics of the system. 
We can see that the HCP-PINN is in good agreement with the ground truth (figure \ref{51_qualitative}a), which is reflected in the low test data error we saw in figure \ref{52_quantative}b, and the mostly white error plot in figure \ref{51_qualitative}d. 
The PINN prediction however struggles to capture the location of the sharp gradient forming at $x=0$ 
(figures \ref{51_qualitative}c, \ref{51_qualitative}e).
It has been shown that the choice of sampling strategy for physics collocation points is an influential hyperparameter in the training of PINNs\cite{florido_2024_pc_sampling}, and there is a sparse region around ($t=0.55,x=0$).
Neither model has access to internal solution data, or physics information in the sparse regions. Despite this, the HCP-PINN generates a more realistic solution in this blind region, evidence that the HCP-PINN is more robust to the sampling choice of physics collocation points.
We have also found that the HCP-PINN seems to require fewer physics collocation points than a vanilla PINN. In further investigations (not included here) we found that a PINN required $M\approx1500$ 
for a comparable test PE to the HCP-PINN with $M=500$, a promising result when the calculation of the residual is a computational bottle neck for both models.

\noindent
The hard constraint projection method introduces two types of user defined hyperparameters to the hyperparameter tuning process.
The first is the formulation of the two matrices $A$ and $B$. Even with the order of the finite discretisation method fixed, the decomposition is non-unique (for example, a constant value and its reciprocal can be used to scale $A$ and $B$). From our preliminary explorations, we have found that this implementation choice has a non-trivial effect on the model's training and performance.
Higher order finite-difference discretisations require more values for the construction of $B_i$ (the discretisation stencil) and as the back-propagation must pass through all elements, the computational expense increases with discretisation order. We intend to investigate if higher order discretisations offer an improvement in performance to offset the computational expense, however this is left for future work.

\noindent
The second HCP specific hyperparameter is the size of the discretisation, $\Delta t, \Delta x$ in equation (\ref{burgers_discA}). 
Decreasing the values of $\Delta t$ and $\Delta x$ has no effect on the computational expense of the model, and analysis into this hyperparameter is ongoing. We have observed that for unrealistic values of $\Delta t, \Delta x\geq 1.0$ the model performs poorly, and that models where $\Delta t > \Delta x$ seem to exhibit higher accuracy (lower test data error) than for $\Delta t < \Delta x$.

\noindent
When Chen \textit{et al.} \cite{chen2021HCP} applied the hard constraint projection method to a linear PDE they reported that all components of the loss declined faster than for the comparable PINN, but that with enough training the PINN converged to the same value.
We speculate that this discrepancy is due to the nonlinear nature of the Burgers' PDE, but we cannot yet discount the influence of PDE parameters (the viscosity of the system) or the dimension of the PDE being solved.
We have also started to implement the HCP method in a PINN that predicts solutions to the 2D incompressible Navier Stokes equations.
The initial results from the Navier Stokes HCP-PINN have been less favourable than for the Burgers' HCP-PINN. There is evidence that when predicting solutions to the Navier Stokes equations, that the HCP-PINN predictions can be trapped in local minima that the vanilla PINN is able to escape from, and the results we have from our hyperparameter tests are so far inconclusive.
We expect that upon the completion of our investigations into the 1D Burgers' HCP-PINN we will be better informed when revisiting the method implementation for higher dimensions and more complex PDEs.





\section{Conclusion}
We present preliminary results on the HCP-PINN, a PINN with an unlearnable hard constraint projection layer, tested on the 1D Burgers equation, a nonlinear convection-diffusion PDE.
The training data comprised of solution data only at the spatial and temporal boundaries, and physics collocation points (on which the governing equation is weakly enforced) through the domain interior. 

\noindent
The HCP-PINN and equivalent PINN trained similarly for an initial training period, after which the PINN with no HCP appeared to plateau, and the HCP-PINN continued to decline until it converged to a lower value than for the PINN. 
When tested on a grid of points covering the entire space-time domain, the HCP-PINN exhibits higher accuracy than the comparable PINN, converging to a data error roughly one order of magnitude lower. 
We also observe that the inclusion of the HCP layer results in better observance of the governing PDE.
Our findings suggest that the HCP-PINN is more robust to the selection of physics collocation points, and less dependent on collocation point quantity.

\noindent
The authors continue investigations into HCP specific hyperparameters, alongside planning to explore other PDE parameter regimes, higher dimensional Burgers' equations, and other nonlinear PDEs such as the incompressible Navier Stokes equations.

\fontsize{9}{9}\selectfont 

\bibliographystyle{ieeetr}
\bibliography{references2}

@article{brunton2020rev,
author = {Brunton, Steven L. and Noack, Bernd R. and Koumoutsakos, Petros},
title = {Machine Learning for Fluid Mechanics},
journal = {Annual Review of Fluid Mechanics},
volume = {52},
number = {1},
pages = {477 - 508},
year = {2020},
doi = {10.1146/annurev-fluid-010719-060214},
eprint = { https://doi.org/10.1146/annurev-fluid-010719-060214}
}

@inproceedings{kingma2017adam,
    author={Diederik P. Kingma and Jimmy Lei Ba},
    editor       = {Yoshua Bengio and
                  Yann LeCun},
    title        = {Adam: {A} Method for Stochastic Optimization},
    booktitle    = {3rd International Conference on Learning Representations, {ICLR} 2015,
                  San Diego, CA, USA, May 7-9, 2015, Conference Track Proceedings},
    year         = {2015},
    url          = {http://arxiv.org/abs/1412.6980}
}

@article{chen2021HCP,
	doi = {10.1016/j.jcp.2021.110624},
	year = 2021,
	publisher = {Elsevier {BV}},
	volume = {445},
	pages = {110624},
	author = {Yuntian Chen and Dou Huang and Dongxiao Zhang and Junsheng Zeng and Nanzhe Wang and Haoran Zhang and Jinyue Yan},
	title = {Theory-guided hard constraint projection ({HCP}): A knowledge-based data-driven scientific machine learning method},
	journal = {Journal of Computational Physics}
}

@article{raissi2019ogPINN,
	title = {Physics-informed neural networks: {A} deep learning framework for solving forward and inverse problems involving nonlinear partial differential equations},
	volume = {378},
	issn = {0021-9991},
	doi = {https://doi.org/10.1016/j.jcp.2018.10.045},
	journal = {Journal of Computational Physics},
	author = {Raissi, M. and Perdikaris, P. and Karniadakis, G. E.},
	year = {2019},
	pages = {686 - 707},
}

@Article{tong2020gappypod,
    AUTHOR = {Tong, Zheming and Li, Yue},
    TITLE = {Real-Time Reconstruction of Contaminant Dispersion from Sparse Sensor Observations with Gappy POD Method},
    JOURNAL = {Energies},
    VOLUME = {13},
    YEAR = {2020},
    NUMBER = {8},
    ARTICLE-NUMBER = {1956},
    URL = {https://www.mdpi.com/1996-1073/13/8/1956},
    ISSN = {1996-1073},
    DOI = {10.3390/en13081956}
}

@article{nakamura2021CNNAELSTM,
	title = {Convolutional neural network and long short-term memory based reduced order surrogate for minimal turbulent channel flow},
	volume = {33},
	doi = {10.1063/5.0039845},
	number = {2},
	journal = {Physics of Fluids},
	author = {Nakamura, Taichi and Fukami, Kai and Hasegawa, Kazuto and Nabae, Yusuke and Fukagata, Koji},
	year = {2021},
	note = {Publisher: AIP Publishing},
	pages = {025116},
}

@article{mohan2020PhyCAE,
    author = {Mohan, Arvind and Lubbers, Nicholas and Chertkov, Misha and Livescu, Daniel},
    year = {2023},
    month = {01},
    pages = {014604},
    numpages = {17},
    title = {Embedding hard physical constraints in neural network coarse-graining of three-dimensional turbulence},
    volume = {8},
    journal = {Physical Review Fluids},
    doi = {10.1103/PhysRevFluids.8.014604}
}

@article{cai2021espresso, 
    title={Flow over an espresso cup: inferring 3-D velocity and pressure fields from tomographic background oriented Schlieren via physics-informed neural networks}, 
    volume={915}, 
    DOI={10.1017/jfm.2021.135}, 
    journal={Journal of Fluid Mechanics}, 
    publisher={Cambridge University Press}, 
    author={Cai, Shengze and Wang, Zhicheng and Fuest, Frederik and Jeon, Young Jin and Gray, Callum and Karniadakis, George Em}, 
    year={2021}, 
    pages={A102}
}

@article{bajaj2021G-PINN,
	doi = {10.1088/2632-2153/acb416},
	url = {https://doi.org/10.1088%2F2632-2153%2Facb416},
	year = 2021,
	publisher = {{IOP} Publishing},
	volume = {4},
	number = {1},
	pages = {015013},
	author = {Chandrajit Bajaj and Luke McLennan and Timothy Andeen and Avik Roy},
	title = {Recipes for when physics fails: recovering robust learning of physics informed neural networks},
	journal = {Machine Learning: Science and Technology}
}

@article{lu2021hardcon_PINN_inverse,
    author = {Lu, Lu and Pestourie, Rapha\"{e}l and Yao, Wenjie and Wang, Zhicheng and Verdugo, Francesc and Johnson, Steven G.},
    title = {Physics-Informed Neural Networks with Hard Constraints for Inverse Design},
    journal = {SIAM Journal on Scientific Computing},
    volume = {43},
    number = {6},
    pages = {B1105-B1132},
    year = {2021},
    doi = {10.1137/21M1397908}
}

@inproceedings{negiar2023hardconst,
  author       = {Geoffrey N{\'{e}}giar and
                  Michael W. Mahoney and
                  Aditi S. Krishnapriyan},
  title        = {Learning differentiable solvers for systems with hard constraints},
  booktitle    = {The Eleventh International Conference on Learning Representations,
                  {ICLR} 2023, Kigali, Rwanda, May 1-5, 2023},
  publisher    = {OpenReview.net},
  year         = {2023}
}

@inproceedings{krishnapriyan2021PINNflaws,
     author = {Krishnapriyan, Aditi and Gholami, Amir and Zhe, Shandian and Kirby, Robert and Mahoney, Michael W},
     booktitle = {Advances in Neural Information Processing Systems},
     editor = {M. Ranzato and A. Beygelzimer and Y. Dauphin and P.S. Liang and J. Wortman Vaughan},
     pages = {26548--26560},
     publisher = {Curran Associates, Inc.},
     title = {Characterizing possible failure modes in physics-informed neural networks},
     url = {https://proceedings.neurips.cc/paper_files/paper/2021/file/df438e5206f31600e6ae4af72f2725f1-Paper.pdf},
     volume = {34},
     year = {2021}
}

@article{BEUCLER_TOM2021analconst,
	title = {Enforcing {Analytic} {Constraints} in {Neural} {Networks} {Emulating} {Physical} {Systems}},
	volume = {126},
	doi = {10.1103/physrevlett.126.098302},
	number = {9},
	journal = {Physical Review Letters},
	author = {Beucler, Tom and Pritchard, Michael and Rasp, Stephan and Ott, Jordan and Baldi, Pierre and Gentine, Pierre},
	year = {2021},
	note = {Publisher: American Physical Society (APS)},
}

@inproceedings{florido_2024_pc_sampling,
author = {Florido, Jose and Wang, He and Khan, Amirul and Jimack, Peter K.},
title = {Investigating Guiding Information for Adaptive Collocation Point Sampling in PINNs},
year = {2024},
isbn = {978-3-031-63758-2},
publisher = {Springer-Verlag},
address = {Berlin, Heidelberg},
url = {https://doi.org/10.1007/978-3-031-63759-9_36},
doi = {10.1007/978-3-031-63759-9_36},
booktitle = {Computational Science – ICCS 2024: 24th International Conference, Malaga, Spain, July 2–4, 2024, Proceedings, Part III},
pages = {323–337},
numpages = {15}
}

@InProceedings{shone2023cardiacmriPINN,
author="Shone, Fergus
and Ravikumar, Nishant
and Lassila, Toni
and MacRaild, Michael
and Wang, Yongxing
and Taylor, Zeike A.
and Jimack, Peter
and Dall'Armellina, Erica
and Frangi, Alejandro F.",
editor="Frangi, Alejandro
and de Bruijne, Marleen
and Wassermann, Demian
and Navab, Nassir",
title="Deep Physics-Informed Super-Resolution of Cardiac 4D-Flow MRI",
booktitle="Information Processing in Medical Imaging",
year="2023",
publisher="Springer Nature Switzerland",
pages="511--522",
isbn="978-3-031-34048-2"
}


\end{document}